\documentclass[letter,twocolumn]{jpsj3}
\usepackage{txfonts}
\usepackage{color}

\title{Interplay between quadrupolar and magnetic interactions in 5$d^{1}$ double perovskite Ba$_2$MgReO$_6$ under pressure}

\author{
Hiroto~Arima$^1$,
Yoshiaki Oshita$^{1}$, 
Daigorou~Hirai$^{2}$, 
Zenji~Hiroi$^{2}$,
and 
Kazuyuki~Matsubayashi$^1$\thanks{k.matsubayashi@uec.ac.jp}
}

\inst{
$^1$Department of Engineering Science, The University of Electro-Communications, Chofu, Tokyo 182-8585, Japan \\ 
$^2$ Institute for Solid State Physics, The University of Tokyo, Kashiwa, Chiba 277-8581, Japan
} 

\abst{
We measured the electrical resistivity, AC magnetic susceptibility, and specific heat of the cubic double perovskite Ba$_2$MgReO$_6$ under pressure. The application of pressure leads to a gradual increase in the transition temperature to the canted antiferromagnetic order, while the quadrupolar order is suppressed above $\sim$5~GPa, at which a collinear antiferromagnetic state appears. The obtained unit cell volume-temperature phase diagram for the compound as well as the related compounds suggests that Ba$_2$MgReO$_6$ exists near the phase boundary between the noncollinear and the collinear antiferromagnetic order with multipole decrees of freedom.
}

\begin{document}
\maketitle

	Spin-orbit coupling and electronic correlation are the key components for the formation of a spin-orbital entangled state, which is characterized by the effective total angular momentum $J$. Exotic states of matter, such as multipolar orders resulting from the combination of spin-orbit coupling and crystal electric fields, have been extensively investigated in many $f$-electron systems \cite{Kuramoto}. Examples include Pr$Tr_{2}X_{20}$ ($Tr$; transition metal, $X$ = Al, Zn), where quadrupolar and possible octupolar orders are realized \cite{Onimaru,Sakai,Patri,Hattori}. In particular, the high degree of tunability of the multipolar phase using magnetic fields and pressure allows for studying the interplay between the multipolar degrees of freedom and other electronic states, including heavy fermion superconductivity \cite{Yoshida,Shimura,Matsubayashi,Umeo}.
	
	Double perovskite oxides with the chemical formula $A_{2}BB^{'}$O$_{6}$ also provide an intriguing basis for various phases with spin-orbit coupling; the interplay between the multipole order and magnetic order is of particular interest in this regard \cite{Chen,Krempa,Takayama}. Transition metal $B^{'}$ ions with the 5$d^{1}$ configuration, such as Os$^{7+}$ and Re$^{6+}$, in a face-centered cubic (fcc) lattice, leads to $J_{\rm eff}$ = 3/2 quartet characterized by multipole degrees of freedom. The relatively large spatial separation of $B^{'}$ ions reduces electron hopping and results in a Mott insulating state. Indeed, the Mott insulating Ba$_{2}$NaOsO$_{6}$ undergoes an unusual noncollinear antiferromagnetic order at $T_{\rm m}$ = 6.8~K with a small saturation moment of 0.2~$\mu_{\rm B}$ \cite{Erikson}, which is driven by the staggered quadrupolar order \cite{Chen,Lu}. A recent calorimetric measurement and the corresponding magnetic field evolution on a single crystal reveal the existence of the phase boundary into the broken local point symmetry phase \cite{Willa}, as observed in a previous NMR experiment \cite{Lu}. Ba$_{2}$MgReO$_{6}$ is another promising material for realizing both exotic magnetic order and quadrupolar order \cite{Marjerrison,Hirai_1}. Ba$_{2}$MgReO$_{6}$ undergoes a magnetic transition at $T_{\rm m}$ = 18~K, exhibiting magnetic properties resembling those of Ba$_{2}$NaOsO$_{6}$. The specific heat of Ba$_{2}$MgReO$_{6}$ exhibits a broad anomaly at $T_{\rm q}$ = 33~K, and the total entropy above $T_{\rm q}$ is close to $R$ln 4; this reflects a quartet state that causes a quadrupolar order. The magnetic order with [110] anisotropy and a small saturation moment of $\sim$0.3 $\mu_{\rm B}$ is unusual according to the Landau theory; additionally, the magnetic structure affected by the preceding quadrupole order is suggested to be composed of two sublattices that spin-align ferromagnetically in the (001) plane, whereas the planes are coupled antiferromagnetically along the [001] direction \cite{Hirai_2}. Because of spin canting, an uncompensated moment along the [110] direction appears below $T_{\rm m}$, which is consistent with predictions based on the mean-field theory \cite{Chen}. This canted antiferromagnetic state, accompanied by the quadrupole order, has recently been found in Ba$_2$CdReO$_6$ \cite{Hirai_3}. However, Ba$_2$CaReO$_6$, which has a larger unit cell volume, exhibits only a collinear antiferromagnetic order without canting \cite{Yamamura}. More interestingly, it is theoretically suggested that the octupolar moment can dominate the dipole moment in the collinear antiferromagnetic phase \cite{Chen}. The emergence of the octupolar order has also been proposed for cubic 5$d^{2}$ double perovskites Ba$_2$$B$OsO$_6$ ($B$ = Mg, Zn, Ca) on the fcc lattice \cite{Paramekanti,Lovesey,Maharaj}. However, experimental verification of the octupolar order remains a major challenge. 
	
	A better understanding of the complex interplay between the multipole and magnetic order requires the construction of a phase diagram through experiments. External pressure, as a control parameter, can continuously drive the system into other intriguing phases, however, few studies on double perovskite oxides with the multipole degrees of freedom has been made so far. In this study, we measured the electrical resistivity, AC magnetic susceptibility, and specific heat of a Ba$_{2}$MgReO$_{6}$ single crystal under pressure. Moreover, we obtained the global phase diagram together with the phase diagrams of the sister compounds Ba$_2$$B$ReO$_6$ ($B$ = Zn, Cd, Ca).
	
	Single crystals of Ba$_{2}$MgReO$_{6}$ were grown by the flux method \cite{Hirai_1}. The electrical resistivity was measured using a standard four-probe technique with the current flow along the $\left[110\right]$ axis. The AC magnetic susceptibility was measured at a fixed frequency of 231 Hz with a modulation field of 0.4 mT applied along the $\left[110\right]$ axis. AC calorimetric measurements were performed to determine the relative value of the specific heat. Specifically, an Au/AuFe thermocouple and strain gauge were used as the thermometer and heater, respectively. The heater power $P$ modulated by the frequency $\omega$/2 of the AC current produces temperature oscillations of the sample at a frequency $\omega$ \cite{Matsubayashi_spe}. An opposed-anvil-type high-pressure cell was used to generate hydrostatic pressure \cite{Kitagawa}, and glycerol was used as the pressure-transmitting medium (PTM). The pressure at low temperature was calibrated by monitoring the pressure dependence of the superconducting transition temperature of the lead placed beside the Ba$_{2}$MgReO$_{6}$ single crystal inside the sample chamber.

\begin{figure}[t]
\begin{center}
\includegraphics[width=0.5\textwidth]{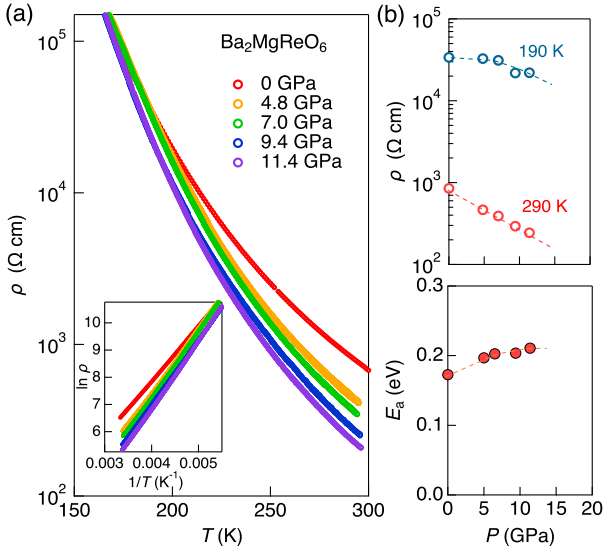}
\end{center}
\caption{(Color online) (a) Temperature dependence of the electrical resistivity $\rho(T)$ of Ba$_2$MgReO$_6$ at selected pressures. The inset shows ln $\rho$ as a function of the inverse temperature. (b) Pressure dependence of electrical resistivity at 190 and 290 K (upper panel) and activation energy (lower panel). The lines are visual guides.}
\label{Fig1}
\end{figure}

	Figure 1(a) shows the temperature dependence of the electrical resistivity  $\rho$ under high pressure for Ba$_2$MgReO$_6$. The insulating behavior at ambient pressure with an activation energy $E_{\textrm a}$ $\sim$ 0.16~eV is consistent with that mentioned in a previous report \cite{Hirai_1}. In the present study, the activation energy was estimated from the slope of the Arrhenius plot using the following equation: $\rho =\rho_0 \exp\left(E_\textrm{a}/k_\textrm{B}T\right)$, where $E_{\textrm a}$ is the activation energy, and $k_{\textrm B}$ is the Boltzmann constant. With increasing pressure, the overall resistivity in the high-temperature region decreases in magnitude; however, the pressure variation of the resistivity at low temperatures is very weak up to the highest pressure investigated, as evident from plotting the resistivity values at a fixed temperature as a function of pressure (see Fig.~1(b)). The exponential decrease in the resistivity at room temperature is attributable to the suppression of the energy gap; however, the energy gap evaluated from the temperature dependence increases slightly with pressure owing to the persistence of the high resistivity value at low temperature. Note that the activation-type $T$-dependence over the entire temperature range is maintained up to $\sim$11~GPa, as shown in the inset of Fig.~1(a). The origin of the pressure response for the energy gap is unclear, but the Mott insulating state is expected to be preserved at low temperatures, together with the $J_{\rm eff}$ = 3/2 quartet ground state with multiple degrees of freedom.

\begin{figure}[t]
\begin{center}
\includegraphics[width=0.45\textwidth]{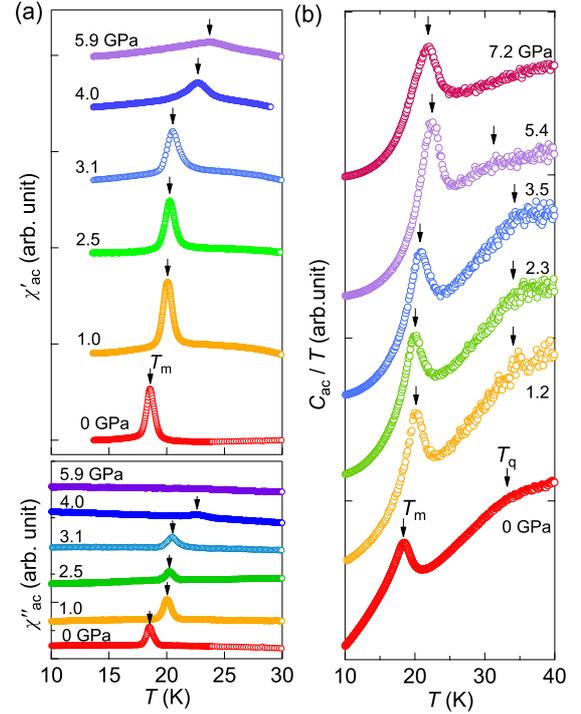}
\end{center}
\caption{(Color online) Temperature dependence of (a) the real (upper panel) and imaginary (lower panel) parts of the AC magnetic susceptibility $\chi_{\rm ac}^{'}$ and $\chi_{\rm ac}^{''}$, respectively, and (b) the specific heat of Ba$_2$MgReO$_6$ under pressure. The arrows at $T_{\rm m}$ and $T_{\rm q}$ indicate the magnetic and quadrupole transition temperatures, respectively. The data for different pressures are shifted for clarity.}
\label{Fig2}
\end{figure}

	Next, we focus on the pressure variation of the quadrupole order and the canted antiferromagnetic order traced via AC magnetic susceptibility and specific heat measurements. Figure~2(a) shows the temperature dependence of the real and imaginary parts of the AC magnetic susceptibility $\chi_{\rm ac}^{'}$ and $\chi_{\rm ac}^{''}$, respectively. At ambient pressure, a distinct peak is observed in both $\chi_{\rm ac}^{'}$ and $\chi_{\rm ac}^{''}$ at $T_{\rm m}$~$\sim$~18~K; this indicates a canted antiferromagnetic transition \cite{Hirai_1}. With increasing pressure, $T_{\rm m}$ gradually shifts to a higher temperature. Remarkably, the signal of $\chi_{\rm ac}^{'}$ at $T_{\rm m}$ decreases significantly at pressures above 4~GPa, accompanied by a further increase in $T_{\rm m}$. In particular, the peak structure in $\chi_{\rm ac}^{''}$ disappears completely at 5.9~GPa. As discussed subsequently, this is likely associated with the change in the magnetic structure. Furthermore, the specific heat measurements reveal not only the magnetic transition at $T_{\rm m}$ but also an anomaly at $T_{\rm q}$ due to the quadrupole order \cite{Hirai_1}. As shown in Fig.~2(b), at ambient pressure, the broad anomaly associated with the quadrupole order is observed at $T_{\rm q}$ $\sim$ 33~K, which is consistent with the previous report \cite{Hirai_1}. Note that a similar broadening of the transition is also observed in the sister compound Ba$_2$CdReO$_6$, in which $J_{\rm eff}$ = 3/2 quartet state with multipole decrees of freedom is realized \cite{Hirai_3}. At pressures below 3.5~GPa, $T_{\rm q}$ is almost pressure-independent; however, with a further increase in the pressure, the anomaly at $T_{\rm q}$ becomes weak and shifts to a lower temperature, and ultimately becomes invisible at $\sim$7.2~GPa.

\begin{figure}[t]
\begin{center}
\includegraphics[width=0.33\textwidth]{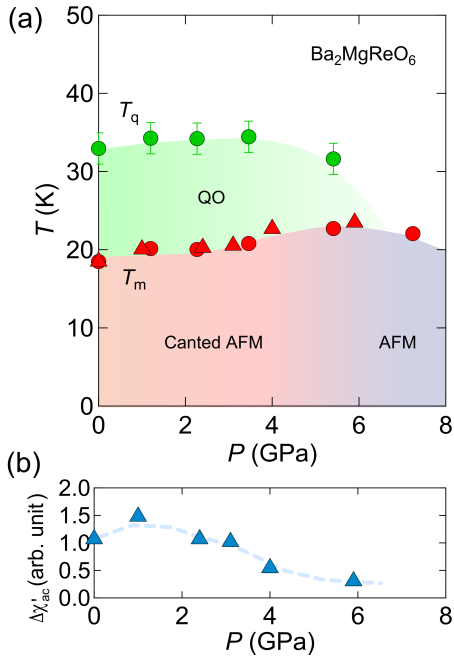}
\end{center}
\caption{
(Color online) (a) Temperature-pressure phase diagram of Ba$_2$MgReO$_6$. The quadrupole ordering (QO) temperatures $T_{\rm q}$ determined from the specific heat data (closed circles). The magnetic transition temperatures $T_{\rm m}$ are deduced from the peak position in the temperature dependence of the specific heat (closed circles) and the AC magnetic susceptibility (closed triangles), respectively. Applying pressure changes the magnetic structure from the canted to the collinear antiferromagnetic (AFM) state. The colors are visual guides. (b) Pressure dependence of the relative magnitude of the AC magnetic susceptibility signal $\Delta\chi_{\rm ac}^{'}$, which is estimated by subtracting the signal at 30 K as a paramagnetic region from that at $T_{\rm m}$. $\Delta\chi_{\rm ac}^{'}$ is displayed in arbitrary units, but the relative magnitude can be compared because the same experimental conditions, such as frequency and modulation field, are identical.
}
\label{Fig3}
\end{figure}

	Figure 3 shows the pressure--temperature phase diagram obtained from the AC magnetic susceptibility and specific heat data. In the low-pressure region, Ba$_2$MgReO$_6$ experiences a quadrupole transition at $T_{\rm q}$, followed by the canted antiferromagnetic transition at $T_{\rm m}$. Intriguingly, these transition temperatures exhibit a trend opposite to that of the pressure dependence, especially above 4 GPa, and seem to merge at approximately $\sim$6~GPa, where the relative magnitude of the real-part signal for AC magnetic susceptibility at $T_{\rm m}$ becomes considerably small (see Fig.~3(b)). This indicates suppression of spontaneous ferromagnetic components. Therefore, we speculate that a gradual increase in the canting angle with pressure alters the magnetic structure to a collinear antiferromagnetic state, as observed in related compounds, such as Ba$_2$CaReO$_6$ and Sr$_2$MgReO$_6$ \cite{Yamamura,Gao}

	Here, the complications arising from the high sensitivity to the pressure conditions in the multipole order must be considered. Indeed, according to a recent high-pressure experiment on PrIr$_2$Zn$_{20}$ \cite{Umeo}, the collapse of the quadrupole order is strongly affected by the hydrostaticity of the PTM; the transition temperature of the quadrupole order at $T_{\textrm q}$ is monotonically increased by applying pressure with an argon PTM, which can maintain a fairly high hydrostaticity up to $\sim$10~GPa. On the other hand, abrupt collapse of $T_{\textrm q}$ with a glycerol PTM is observed above $\sim$5~GPa, where the effect of non-hydrostaticity starts to manifest 	 owing to the solidification of glycerol \cite{Tateiwa}. The use of a glycerol PTM in the present study does not rule out the possibility that the solidification of the liquid pressure medium causes pressure inhomogeneity or uniaxial stress, especially above $\sim$5~GPa, resulting in the suppression of $T_{\rm q}$ because of the lifting of the quadrupolar degeneracy of Ba$_2$MgReO$_6$. However, we claim that the non-hydrostatic effect is not very significant because both the magnetic transition at $T_{\rm m}$ in the specific heat and the superconducting transition of the pressure manometer remain sharp up to the highest pressure investigated in this study. More importantly, the suppression of $T_{\rm q}$ coincides with a change in the magnetic structure from the canted to the collinear antiferromagnetic state, which suggests an intrinsic interplay between the quadrupole and magnetic interactions. This is reminiscent of the effect of pressure on the magnetic order and the quadrupole order in CeB$_{6}$ with the $\Gamma_{8}$ quartet ground state, which is identical to the $J_{\rm eff}$ = 3/2 quartet state in Ba$_2$MgReO$_6$. In CeB$_{6}$, the quadrupole and magnetic order are pressure-sensitive\cite{Sullow, Ikeda}, reflecting a subtle balance between the quadrupole, antiferromagnetic, and ferromagnetic interactions and the relatively small characteristic energy scale in heavy fermion systems.

\begin{figure}[t]
\begin{center}
\includegraphics[width=0.42\textwidth]{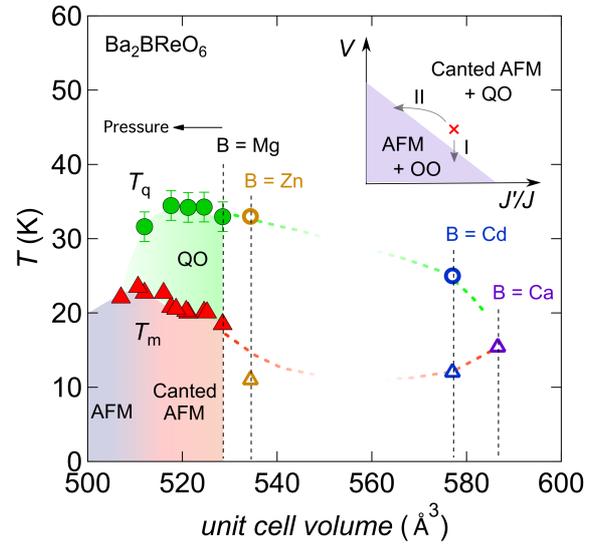}
\end{center}
\caption{(Color online) 
Phase diagram of Ba$_2$$B$ReO$_6$ ($B$ = Mg, Zn, Cd, Ca) as a function of the unit cell volume. $T_{\rm q}$ and  $T_{\rm m}$ are taken from Ref.~17,19 ($B$ = Mg and Cd), Ref.~16 ($B$ = Zn), and Ref.~20 ($B$ = Ca). For Ba$_2$MgReO$_6$, the pressure values are transformed into the unit cell volume using the Birch-Murnaghan equation (see text). The lines and colors are visual guides. The inset shows the schematic mean-field zero-temperature phase diagram proposed in Ref.~10. The red cross in the phase diagram indicates the location of Ba$_2$MgReO$_6$ at ambient pressure. The lines with arrows denote trajectories I and II for the effect of B-site substitution and pressure, respectively. AFM, QO, and OO represent the antiferromagnetic, quadrupole order, and octupole order, respectively.
}
\label{Fig4}
\end{figure}

	To discuss the rich variety of 5$d^{1}$ double perovskite compounds with energy scales larger than those of the $f$-electron system, the phase diagram in a wide unit cell volume--temperature parameter space together with the sister compounds Ba$_2$$B$ReO$_6$ ($B$ = Zn, Cd, Ca) is shown in Fig.~4. We estimated the pressure-volume relationship using the third-order Birch-Murnaghan equation of state. The bulk modulus $B$ and its derivative $B'$ were 158.3 GPa and 4.8, respectively, which were derived from first-principles calculations \cite{Dar}. The substitution of Mg with the larger Zn or Cd ions causes an increase in the cell volume, viewed as a negative pressure on Ba$_2$MgReO$_6$; thus, the transition temperatures $T_{\rm q}$ and $T_{\rm m}$ in Ba$_2$ZnReO$_6$ and Ba$_2$CdReO$_6$ are somewhat consistent with those determined via extrapolation of data from high-pressure experiments on Ba$_2$MgReO$_6$. In contrast, for Ba$_2$CaReO$_6$, which has the largest unit cell volume in this series of compounds, only the magnetic transition at $T_{\rm m}$ survives, and the magnetic structure changes from the canted to the collinear antiferromagnetic state \cite{Yamamura}. As mentioned previously \cite{Hirai_3}, this chemical trend can be understood based on the mean-field theory \cite{Chen} by introducing three interactions: nearest-neighbor AF exchange ($J$), FM exchange ($J^{\prime}$), and quadrupolar interaction ($V$). $V$ can be described using the following equation: $V$ = 9$\sqrt{2}$$Q^2$/$a^5$, where $Q$ and $a$ represent the magnitude of the electric quadrupole and the lattice constant, respectively. Because the volume expansion leads to a decrease in $Q$ due to weaker hybridization between the Re 5$d$ and O 2$p$ states, the substitution from Mg to Ca reduces the value of $V$. Assuming a relatively small value of $J^{\prime}/J  = 0.2$ inferred from the negative Weiss temperature \cite{Marjerrison,Hirai_1,Hirai_3,Yamamura}, the volume evolution of the ground state in Ba$_2$$B$ReO$_6$ ($B$ = Mg, Zn, Cd, Ca) agrees with the theoretical predictions, which correspond to trajectory I in the schematic mean-field zero-temperature phase diagram, as shown in the inset of Fig.~4. The mean-field calculation also predicts the quadrupole transition temperature as follows and can explain the suppression of $T_{\rm q}$ under the decrease in $V$, resulting in a constant value of $J^{\prime}/J $:  
\begin{equation}
T_{\rm q} = \frac{43V+18J^{\prime}-3J}{18}.
\end{equation}
However, the unit cell volume dependence of $T_{\rm q}$ tends to saturate in the regime of data obtained from the high-pressure experiment, and further decreasing volume, $T_{\rm q}$ starts to be suppressed. By considering the monotonic increase in $V$ with decreasing cell volume, these results can be ascribed to the reduction in $J^{\prime}/J$, which is schematically shown in the inset of Fig.~4 for trajectory II. This is consistent with the occurrence of the antiferromagnetic ordered phase in the small-volume regime. Furthermore, this interpretation is supported by the following observation in Sr$_2$MgReO$_6$, which has a smaller cell volume because of the smaller ionic radius of Sr with respect to Ba: the negative Weiss temperature ($\Theta_{\rm CW}$ = -134~K) is enhanced compared with that in Ba$_2$$B$ReO$_6$ ($B$ = Mg, Zn, Cd, Ca), and an antiferromagnetic order occurs at $\sim$55~K\cite{Gao}. All these results strongly suggest that Ba$_2$$B$ReO$_6$ is located near the phase boundary between the canted and collinear antiferromagnetic orders, which might be dominated by the hidden nontrivial octupole order, as predicted theoretically \cite{Chen}. Further high-pressure studies using techniques such as high-resolution X-ray diffraction or more precise thermodynamic measurements are needed to clarify the nature of exotic phases with multipolar degrees of freedom.

	In summary, we measured the electrical resistivity, AC magnetic susceptibility, and specific heat of Ba$_2$MgReO$_6$ single crystals under pressure and found the suppression of the quadrupole order accompanied by the change in the magnetic structure from the noncollinear to the collinear antiferromagnetic state. The obtained phase diagram, together with the chemical substitution, over a wide range of unit cell volumes provides an intriguing route to uncover the interplay between the multipolar and magnetic order in a series of Ba$_2$$B$ReO$_6$ ($B$ = Mg, Zn, Cd, and Ca) compounds with the 5$d^1$ electronic configuration. \\

\begin{acknowledgment}
	We thank Y.~Uwatoko and K.~Kitagawa for experimental support of high-pressure cell and discussions. This work was partially supported by JSPS KAKENHI Grant Numbers 18H01172, 18H04312(J-Physics), 19H01836, 20H01858, and 21K03442. The use of the facilities at the Coordinated Center for UEC Research Facilities and the Cryogenic Center at the University of Electro-Communications is appreciated.\end{acknowledgment}

\bibliography{Reference}

\end{document}